\documentclass[12pt,letterpaper]{article}
\usepackage[usenames, dvipsnames, table]{xcolor}
\usepackage{jheppub}

\usepackage{tocloft}
\usepackage[]{todonotes}
\usepackage{bbold}
\usepackage{graphicx}
\usepackage{verbatim}
\usepackage[mathscr]{euscript}
\usepackage{slashed}
\usepackage{graphicx}
\usepackage{mathdots}
\usepackage{caption}

\usepackage[labelformat=simple]{subcaption}

\usepackage{enumerate}
\usepackage{bbm}
\usepackage{psfrag}
\usepackage{subfiles}
\usepackage{psfrag}
\usepackage{relsize}
\usepackage{pgfplots}
\usepackage{tikzscale}

\usepackage{bbm} 					
\usepackage{slashed} 				
\usepackage{graphicx}				
\usepackage{subcaption}			
\usepackage{psfrag}				
\usepackage{tensor}				
\usepackage{fouridx}				
\usepackage{bm}					
\usepackage{mdframed}				
\usepackage{multirow}				
\usepackage{soul}					
\usepackage{bbold}				
\usepackage{multicol}				

\usepackage{amsmath}
\usepackage{amssymb}
\usepackage{amsthm}

\usepackage{mathtools}

\usepackage{feynmf}
\usepackage{marvosym}

\usepackage{import}

\newtheoremstyle{named}{}{}{\itshape}{}{\bfseries}{.}{.5em}{#3}
\theoremstyle{named}

\usepackage{diagbox}

\newcommand\mathdiagbox[3][]{\hbox{\tabcolsep=\arraycolsep\diagbox[#1]{$#2$}{$#3$}}}

\newcommand*\xoverline[2][0.75]{%
    \sbox{\myboxA}{$\m@th#2$}%
    \setbox\myboxB\null
    \ht\myboxB=\ht\myboxA%
    \dp\myboxB=\dp\myboxA%
    \wd\myboxB=#1\wd\myboxA
    \sbox\myboxB{$\m@th\overline{\copy\myboxB}$}
    \setlength\mylenA{\the\wd\myboxA}
    \addtolength\mylenA{-\the\wd\myboxB}%
    \ifdim\wd\myboxB<\wd\myboxA%
       \rlap{\hskip 0.5\mylenA\usebox\myboxB}{\usebox\myboxA}%
    \else
        \hskip -0.5\mylenA\rlap{\usebox\myboxA}{\hskip 0.5\mylenA\usebox\myboxB}%
    \fi}
\makeatother

\newcommand{\cF}{\mathcal{F}}

\newcommand{\mKK}{m_{\text{KK}}}
\newcommand{\Ms}{M_{\text{string}}}

\definecolor{cobalt}{RGB}{44, 98, 120}
\definecolor{celadon}{rgb}{0.67, 0.88, 0.69}
\definecolor{dm}{cmyk}{.20, 0, .30, 0}
\definecolor{burgundy}{rgb}{0.5, 0.0, 0.13}
\definecolor{plotBlue}{RGB}{94, 130, 181}

\newcommand{\rmd}{\textrm{d}}
\def\be{\begin{equation}}
\def\ee{\end{equation}}
\def\bea{\begin{eqnarray}}
\def\eea{\end{eqnarray}}

\hypersetup{
  colorlinks,
  citecolor=Violet,
  linkcolor=cobalt,
  urlcolor=Blue}

\newif\iffastcompile

\fastcompilefalse

\iffastcompile
\newcommand{\js}[1]{}
\newcommand{\jsi}[1]{}
\newcommand{\cl}[1]{}
\newcommand{\lm}[1]{}
\else
\newcommand{\js}[1]{\todo[color=cobalt!30,size=\scriptsize, bordercolor=cobalt!30]{JS: #1}}
\newcommand{\jsi}[1]{\todo[color=cobalt!30,size=\scriptsize, bordercolor=cobalt!30, inline]{JS: #1}}
\newcommand{\cl}[1]{\todo[color=burgundy!30, size=\scriptsize, bordercolor=burgundy!30]{CL: #1}}
\newcommand{\lm}[1]{\todo[color=dm!90, size=\scriptsize, bordercolor=dm!90]{LM: #1}}
\fi

\ProvideTextCommandDefault{\Dbar}{%
\leavevmode\lower.5ex\rlap{\hskip-.07em\accent"16}D%
}

\begin{document}
	\newcommand{\main}{.}
\begin{titlepage}

\setcounter{page}{1} \baselineskip=15.5pt \thispagestyle{empty}

\bigskip\

\vspace{1.05cm}
\begin{center}
{\fontsize{20}{28} \bfseries Gopakumar-Vafa Invariants   and the \\ \vspace{0.3cm} Emergent String Conjecture}

 \end{center}
\vspace{1cm}

\begin{center}
\scalebox{0.95}[0.95]{{\fontsize{14}{30}\selectfont Tom Rudelius$^{1,2}$}}
\end{center}

\begin{center}
\vspace{0.25 cm}
\textsl{$^1$Department of Mathematical Sciences, Durham University, Durham DH1 3LE United Kingdom}\\
\textsl{$^2$Department of Physics, University of California, Berkeley, CA 94720 USA}\\

\vspace{0.25cm}

\end{center}

\vspace{1cm}
\noindent 

The Emergent String Conjecture of Lee, Lerche, and Weigand holds that every infinite-distance limit in the moduli space of a quantum gravity represents either a decompactification limit or an emergent string limit in some duality frame. Within the context of 5d supergravities coming from M-theory compactifications on Calabi-Yau threefolds, we find evidence for this conjecture by studying (a) the gauge couplings and (b) the BPS spectrum, which is encoded in the Gopakumar-Vafa invariants of the threefold. In the process, we disuss a testable geometric consequence of the Emergent String Conjecture, and we verify that it is satisfied in all complete intersection Calabi-Yau threefolds in products of projective spaces (CICYs).

 \vspace{1.1cm}

\bigskip
\noindent\today

\end{titlepage}
\setcounter{tocdepth}{2}
\tableofcontents

\section{Introduction}\label{INTRO}

The study of the Landscape and the Swampland of quantum gravity has seen a surge of excitement in recent years. Among the plethora of new ideas, the Emergent String Conjecture \cite{Lee:2019wij} has emerged as one of the most exciting. This conjecture holds that any infinite-distance limit in the moduli space of a quantum gravity theory is either a decompactification limit (in which extra compact dimensions grow to infinite size) or else it is an emergent string limit (in which a fundamental string becomes tensionless).

Compelling evidence for the Emergent String Conjecture has been given in various contexts in string theory \cite{Lee:2019xtm,Lee:2019wij,Baume:2019sry,Xu:2020nlh, Basile:2022zee,  Etheredge:2022opl, Blumenhagen:2023yws}. However, to date there has been little progress towards a bottom-up argument for the Emergent String Conjecture, as there is for other Swampland conjectures like the absence of global symmetries. To this end, the first part of this paper, \S\ref{BOTTOMUP}, provides evidence for the Emergent String Conjecture in 5d supergravity by studying the scaling of gauge couplings in infinite-distance limits, expanding on previous work in \cite{Etheredge:2022opl}. Notably, this argument does not rely on any UV input from string/M-theory, and instead it depends solely on the cubic structure of the prepotential in five dimensions. 

In the second part of this paper, \S\ref{GV}, we provide an additional top-down argument for the Emergent String Conjecture by studying the spectrum of BPS states that become light in infinite-distance limits in 5d supergravity. For M-theory compactified on a Calabi-Yau threefold, this BPS spectrum is encoded in the Gopakumar-Vafa invariants of the threefold, which can be computed in a large class of examples using the methods of \cite{Hosono:1993qy,Hosono:1994ax}. In a decompactification limit to six dimensions, we expect a light tower of BPS Kaluza-Klein modes, each of which is labeled by a Kaluza-Klein momentum $q \in \mathbb{Z}$, which does not grow indefinitely as $q$ increases. In contrast, an emergent string limit features a tower of string oscillator modes, which grows exponentially with increasing mass. By studying how the Gopakumar-Vafa invariants $n_q$ scale with increasing $q$, therefore, we find evidence as to whether the infinite-distance limit in question represents a decompactification limit or an emergent string limit.   

Comparing this with the analysis of \S\ref{BOTTOMUP} for several simple Calabi-Yau threefolds, we find perfect agreement: the GV invariants scale like a Kaluza-Klein tower precisely when the gauge couplings scale like a decompactification limit, and the GV invariants scale like a tower of string oscillator modes precisely when the gauge couplings scale like an emergent string limit. Taken together, this gives strong evidence that these limits are indeed decompactification limits and emergent string limits in appropriate duality frames, consistent with the Emergent String Conjecture. 

Our analysis also yields a surprising geometric implication of the Emergent String Conjecture for Calabi-Yau manifolds, which can be tested in examples through the computation of Gopakumar-Vafa invariants and (classical) triple intersection numbers. In \S\ref{SWEEP}, we carry out these tests in 7820 examples of Calabi-Yau threefolds, extending the analysis of \S\ref{GV} to the entire set of Calabi-Yau threefolds which arise as (favorable) complete intersections of hypersurfaces in products of projective spaces (CICYs). For all such CICYs, we again find perfect agreement between the scaling of the GV invariants and the scaling of the gauge couplings, strongly suggesting that each infinite-distance limit in the vector multiplet moduli spaces of these theories represents either an emergent string limit or a decompactification limit. This is remarkable because all of the limits we consider--even the decompactification limits--arise from shrinking one of the $\mathbb{P}^n$ factors of the ambient space. It is a surprising prediction of the Emergent String Conjecture that shrinking this cycle actually corresponds to decompactifying a circle in a different duality frame, but our analysis suggests that this is indeed the case.

The results of \S\ref{GV}-\ref{SWEEP} were previously anticipated in the works \cite{Lee:2019wij, Cota:2022maf}, which explained how decompactification limits and emergent string limits correspond respectively to limits in which a genus-one fiber or a complex 2-dimensional surface inside the Calabi-Yau threefold shrinks to zero size. Our results align nicely with the results of those works. 

Before we get to the crux of the matter, we first review relevant aspects of 5d supergravity, Calabi-Yau geometry, and Kaluza-Klein compactifications from six to five dimensions; this is covered in the following section.

\section{Review of Supergravity, Calabi-Yau Geometry, and Dimensional Reduction}

In this section, we first review supergravity theories in five dimensions and their construction through compactification of M-theory on a Calabi-Yau threefold. We then review the towers of light states that appear when such a theory is decompactified to six dimensions by taking one of the extra dimensions to be large.

\subsection{5d Supergravity and Calabi-Yau Threefolds}

Many features of a 5d supergravity theory are captured by its prepotential, a cubic homogeneous polynomial:
\begin{equation}
\mathcal{F} = \frac{1}{6} C_{I J K} Y^I Y^J Y^K.
\end{equation}
In an M-theory compactification to 5d on a Calabi-Yau threefold $X$, indices $I,J,K$ run from $1$ to $h^{1,1}(X)$, the constants $C_{IJK}$ are the triple intersection numbers of the manifold, and the moduli $Y^I$ are volumes of certain two-cycles.

The prepotential is subject to the constraint $\cF = 1$, which follows from the fact that the overall volume of the Calabi-Yau is not a vector multiplet modulus in 5d, so the vector multiplet moduli space has dimension $h^{1,1}(X)-1$. In this work, however, we will drop this constraint by instead taking the $Y^I$ to be homogeneous coordinates, which are identified under simultaneous rescaling, $Y^I \sim \lambda Y^I$, $\lambda > 0$. Of course, this is simply a choice of convention, which does not affect our results in any way. 

At a generic point in moduli space, the gauge group is $U(1)^{h^{1,1}(X)}$, and the gauge kinetic matrix is given by
\begin{equation}
  a_{I J} = \frac{\mathcal{F}_I \mathcal{F}_J}{\cF^{4/3}} - \frac{\mathcal{F}_{I J}}{\cF^{1/3}} ,
  \label{eq:gaugekinetic}
\end{equation}
with
\begin{equation}
   \mathcal{F}_I = \partial_I \cF= \frac{1}{2} C_{I J K} Y^J Y^K , 
   \qquad \mathcal{F}_{I
   J} = \partial_I \partial_J \cF  = C_{I J K} Y^K .
   \label{eq:prepottrip}
   \end{equation}
The eigenvalues of the gauge kinetic matrix correspond to the inverse-squares of gauge couplings, $\lambda_I \sim 1/g_I^2$. This means that an eigenvalue of $a_{IJ}$ diverges precisely when a gauge coupling vanishes. Conversely, a gauge coupling diverges precisely when an eigenvalue of $a_{IJ}$ vanishes. By electromagnetic duality, this happens precisely when the gauge coupling of a magnetic 2-form gauge field vanishes.

Meanwhile, the metric on moduli space in homogeneous coordinates is given by
\begin{equation}
g_{IJ} = \frac{2}{3} \frac{\cF_I \cF_J}{\mathcal{F}^2} - \frac{\cF_{IJ}}{\cF} .
\label{hgmet}
\end{equation}
This metric is positive-semidefinite: all eigenvalues are positive except for one null eigenvalue, which corresponds to rescaling $Y^I \rightarrow \lambda Y^I$.

The BPS bound is given by
\be
m(q_I) \geq  \left( \frac{\sqrt{2} \pi}{\kappa_5}\right)^{1/3} |Z| =  \left( \frac{\sqrt{2} \pi}{\kappa_5}\right)^{1/3} \frac{|q_I Y^I|}{\cF^{1/3}}\,,
\label{BPSbound}
\ee
where $q_I \in \mathbb{Z}$ is the charge of the particle under the $I$th $U(1)$, and $\kappa_5^2 = 1/(8 \pi G)$. Particles that saturate the BPS bound are called BPS particles, and the number of BPS particles of charge $q_I$ is counted by the Gopakumar-Vafa invariant $n_{q_I}$ \cite{Gopakumar:1998jq,Gopakumar:1998ii}.\footnote{More precisely, Gopakumar-Vafa invariants compute an index of the number of BPS particles, so there can be a cancellation between BPS particles of different spins. This will not be important for our purposes, however.}

This concludes our lightning review of 5d supergravity and Calabi-Yau geometry. More details can be found in e.g. \cite{Bergshoeff:2004kh, Alim:2021vhs}.

\subsection{Dimensional Reduction}\label{DIMRED}

Consider an Einstein-dilaton theory coupled to a 2-form gauge field in six dimensions 
\be
S =  \int \rmd^6 x \sqrt{-g} \left( \frac{1}{2\kappa_6^2} {\cal R}_D - \frac{1}{2} (\nabla \phi)^2 - \frac{1}{2 g_0^2} {\rm e}^{2\phi} |H_3|^2 \right)  \,. \label{eq:generalaction}
\ee
An action of this form shows up, for instance, in minimal supergravity coupled to a tensor multiplet. In the limit $\phi \rightarrow \infty$, the 2-form gauge field becomes weakly coupled, with gauge coupling
\begin{equation}
g_{B, 6}^2 = g_0^2 {\rm e}^{-2 \phi} \rightarrow 0\,.
\end{equation}
The Weak Gravity Conjecture for strings \cite{Arkanihamed:2006dz} implies that in this limit, a tensionless string emerges, with tension $T \sim g_{B, 6} / \kappa_6$. This leads to an infinite tower of charged particles beginning at the string scale, $M_{\rm string} = \sqrt{2 \pi T} \sim \sqrt{g_{B, 6} / \kappa_6} \sim \exp (- \phi /2)$.

Next, suppose that we dimensionally reduce this theory on a circle, with dimensional reduction ansatz
\be
ds_6^2 = {\rm e}^{\frac{-\rho(x)}{\sqrt{3}}} d{\hat s}_5^2(x) + {\rm e}^{\sqrt{3} \rho(x)} dy^2,
   \label{eq:dimredansatz}
\ee
This yields a canonically normalized radion field $\rho$ in five dimensions, in addition to the canonically normalized dilaton $\phi$ that descends from six dimensions. With this, the 5-dimensional gauge coupling for the string picks up an exponential dependence on the radion:
\begin{equation}
g_B^2 = (2 \pi R) g_0^2 {\rm e}^{-2 \phi - \frac{1}{\sqrt{3}} \rho} \,.
\end{equation}
This leads to a tower of string oscillator modes beginning at the string scale, $M_{\rm string} = \sqrt{2 \pi T} \sim \sqrt{g_{B} / \kappa_5} \sim \exp( - \frac{1}{2} \phi - \frac{1}{2\sqrt{3}} \rho )$.

Circle reduction also leads to a Kaluza-Klein photon, with gauge coupling
\be
e_{\text{KK}}^2 = \frac{2 \kappa_5^2}{R^2} {\rm e}^{- \frac{ 4 }{ \sqrt{3} } \rho}  \, .
\ee
The tower Weak Gravity Conjecture \cite{Arkanihamed:2006dz, Heidenreich:2015nta, Andriolo:2018lvp} implies that in the decompactification limit $e_{\text{KK}} \rightarrow 0$, there is a tower of particles beginning at the scale $m_{\text{KK}} \sim e_{\text{KK}} / \kappa_5 \sim \exp(- 2 \rho / \sqrt{3})$. Of course, this is nothing other than the Kaluza-Klein scale, and the tower in question is simply a tower of Kaluza-Klein modes.

\begin{figure}
\centering
\includegraphics[width=80mm]{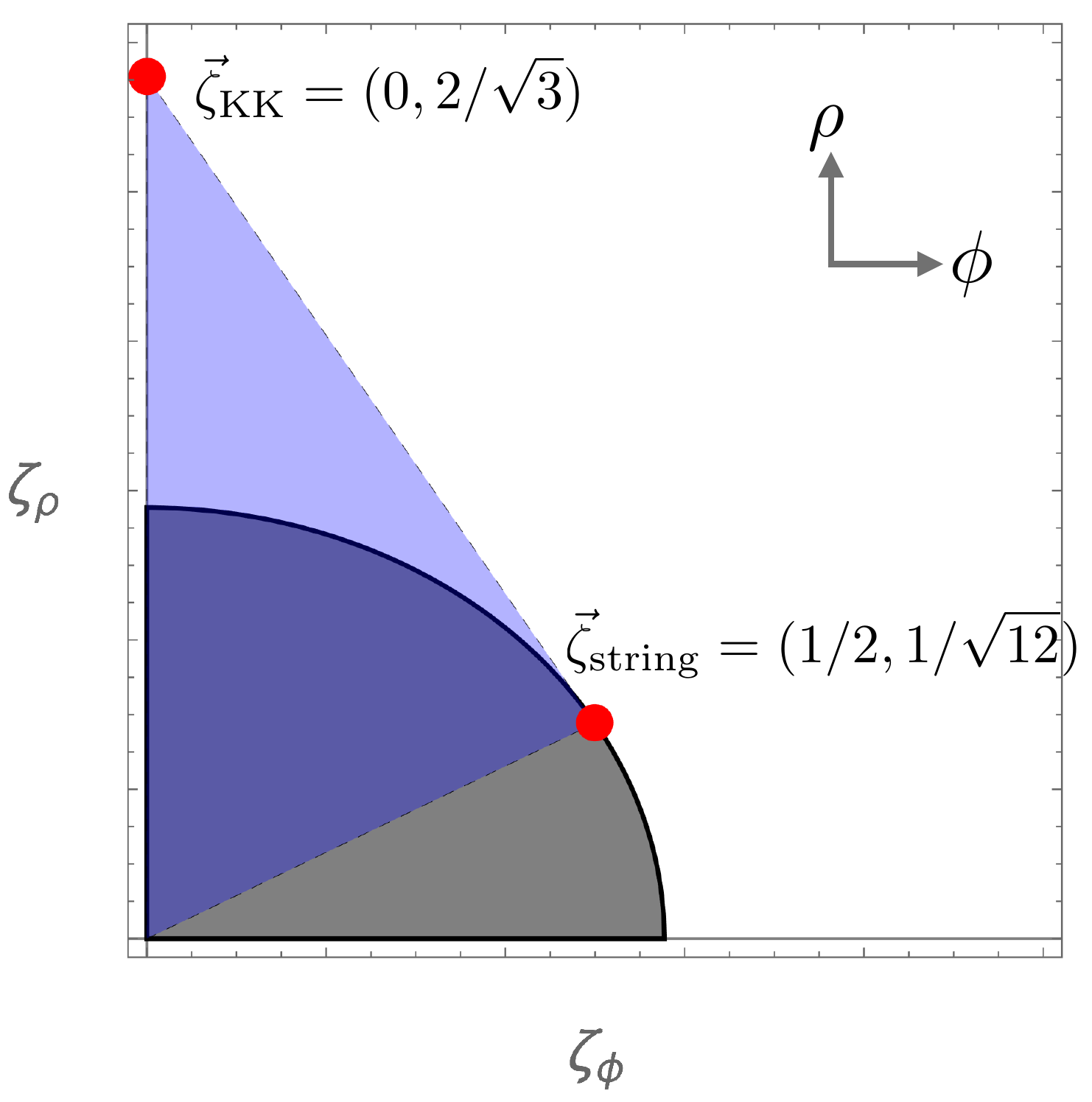}
\caption{The scalar charge-to-mass vectors, $\zeta_i = - \partial_i \log(m)$, for Kaluza-Klein modes and string oscillator modes for a compactification from six to five dimensions. In the string decompactification limit $\rho \rightarrow \infty$, Kaluza-Klein modes and string oscillator modes become light with coefficients $\lambda_{\text{KK}} = 2/\sqrt{3}$ and $\lambda_{\text{string}} = 1/\sqrt{12}$. In the emergent string limit $\rho, \phi \rightarrow \infty$ with $\rmd \rho / \rmd \phi = 1/\sqrt{3}$, Kaluza-Klein modes and string oscillator modes become light with coefficient $\lambda_{\text{KK}} =\lambda_{\text{string}} = 1/\sqrt{3}$. In the intermediate regime (shaded blue), $\rho, \phi \rightarrow \infty$ with $\rmd \rho / \rmd \phi = \tan \vartheta$, Kaluza-Klein modes and string oscillator modes become light with coefficients $  \lambda_{\text{KK}} =  \frac{2}{\sqrt{3}}  \sin(\vartheta) $, $\lambda_{\text{string}}= \frac{1}{2} \cos(\vartheta) + \frac{1}{2\sqrt{3}} \sin(\vartheta)  $.}
\label{thefig}
\end{figure}

It is helpful to define the scalar charge-to-mass vector of a particle of mass $m$ as \cite{Calderon-Infante:2020dhm,Etheredge:2022opl}
\be
\zeta_i \equiv -  \frac\partial{\partial \phi^i}\log m\,,
\label{zetavec}
\ee
where differentiation is performed with the $d$-dimensional Planck mass held fixed. With this, we have
\be
\vec{ \zeta}_{\rm string} = (\frac{1}{2}, \frac{1}{2 \sqrt{3}}  )\,,~~~~\vec{ \zeta}_{\rm string} = (  0 ,  \frac{2}{\sqrt{3}}  ) \,.
\ee
These vectors are plotted in Figure \ref{thefig}.

The Distance Conjecture \cite{Ooguri:2006in} holds that in any infinite-distance limit in moduli space, a tower of particles must become light with a characteristic mass scale that decays exponentially as $m \sim \exp( -  \lambda \kappa_5 \phi  )$. From the discussion above, we see that in the limit $\rho \rightarrow \infty$, the Kaluza-Klein tower satisfies the Distance Conjecture with a coefficient $\lambda = 2/ \sqrt{3}$. We refer to this as the ``strict'' decompactification limit because the radion is taken to infinity while the dilaton is held fixed.
Meanwhile, in the limit $\phi, \rho \rightarrow \infty$ with constant slope $\rmd \rho / \rmd \phi = 1 / \sqrt{3}$, the tower of string oscillator modes satisfies the Distance Conjecture with $\lambda = 1/\sqrt{3}$. This is an emergent string limit, in the terminology of \cite{Lee:2019wij}.
In a more general decompactification limit $\rho, \phi \rightarrow \infty$ with constant slope $\rmd \rho / \rmd \phi \equiv \mathfrak{m}$, let us define a normalized tangent vector $\hat{t} = (\cos \vartheta, \sin \vartheta)$, so that $\tan \vartheta  = \mathfrak{m}$. With this, the Kaluza-Klein modes and the string oscillator modes satisfy the Distance Conjecture in this limit with respective coefficients
\be
\lambda_{\text{KK}} =  \hat{t} \cdot \vec\zeta_{\rm KK} = \frac{2}{\sqrt{3}}  \sin(\vartheta)  \,,~~~~ \lambda_{\text{string}} = \hat{t} \cdot \vec\zeta_{\rm string} = \frac{1}{2} \cos(\vartheta) + \frac{1}{2\sqrt{3}} \sin(\vartheta)     \,.
\label{thetalambda}
\ee
For $\mathfrak{m} \geq 1 / \sqrt{3}$ (shaded blue in Figure \ref{thefig}), these coefficients satisfy $1 / \sqrt{12} \leq \lambda_{\text{string}}  \leq 1 / \sqrt{3} \leq \lambda_{\text{KK}}  \leq 2/\sqrt{3}$.

\section{Decompactification and Emergent Strings}\label{BOTTOMUP}

Consider a path in vector multiplet moduli space of the form $Y^I(s), s \in [0, 1]$, and suppose that the limit $s \rightarrow 0$ lies at infinite distance. 
As shown in \cite{Etheredge:2022opl}, if we restrict the path to take a ``straight-line'' form in homogeneous coordinates, i.e., $Y^I(s) = Y^I_0 + s Y^I_1$, then there are two possible behaviors for the prepotential: either $\cF$ vanishes linearly as $s \rightarrow 0$ ($\cF \sim s$) or else $\cF$ vanishes quadratically as $s \rightarrow 0$ ($\cF \sim s^2$). In the former case, the smallest and largest gauge couplings scale as
\be
g_{\min} \sim \exp(- \frac{2}{\sqrt{3}} \rho )\,,~~~ \frac{1}{\sqrt{g_{\max}}} \sim \exp(- \frac{1}{\sqrt{12}} \rho )\,.
\ee
where $\rho$ is a canonically normalized scalar field which diverges as $\rho \rightarrow \infty$ as $s \rightarrow 0$. Comparing with \S\ref{DIMRED}, we see that this precisely matches the scaling expected for a strict decompactification limit to six dimensions, in which $g_{\min}$ represents the gauge coupling of the Kaluza-Klein photon, $2 \pi/g_{\max}$ represents the gauge coupling for the 2-form gauge field of a fundamental string in six dimensions, and $\rho$ represents the radion field.

The tower Weak Gravity Conjecture (WGC) implies that as a gauge coupling $g$ tends to zero, a tower of light charged particles will appear beginning at the mass scale $m \sim g /\kappa_5$. In this case, setting $g = g_{\min}$, we find a tower of particles whose mass decays in the limit $s \rightarrow 0$ as $\exp(- \lambda \rho)$, with $\lambda = 2/\sqrt{3}$. Meanwhile, the WGC for strings implies that as a 2-form gauge coupling $g_{\text{string}}$ tends to zero, a tower of string oscillator modes will appear at the string scale $M_{\text{string}} \sim \sqrt{g_{\text{string}}} / \kappa_5^{1/2}$. Setting $g_{\text{string}} \sim 1/ g_{\max}$, we find a tower of string oscillator modes whose mass decays in the limit $s \rightarrow 0$ as $\exp(- \lambda \rho)$, with $\lambda = 1/\sqrt{12}$. This precisely matches the expected scaling for a circle compactification of string theory in six dimensions, and it strongly suggests that the limit $\rho \rightarrow \infty$ represents a decompactification limit in some duality frame.

In the second case, $\cF \sim s^2$, the gauge couplings scale as
\be
g_{\min} \sim \exp(- \frac{1}{\sqrt{3}} \phi )\,,~~~ \frac{1}{\sqrt{g_{\max}}} \sim \exp(- \frac{1}{\sqrt{3}} \phi )\,.
\ee
where $\phi$ is a canonically normalized scalar field which diverges as $\phi \rightarrow \infty$ as $s \rightarrow 0$. Comparing with \S\ref{DIMRED}, we see that this precisely matches the expected scaling for an emergent string limit, where.again $g_{\min}$ represents the gauge coupling of a Kaluza-Klein photon and $2 \pi/g_{\max}$ represents the gauge coupling for the 2-form gauge field of a fundamental string.
Invoking the tower WGC and the magnetic WGC for strings, respectively, we find a tower of particles with $\lambda = 1/\sqrt{3}$ and a tower of string oscillator modes with $\lambda = 1/\sqrt{3}$; this precisely matches the scaling expected for an emergent string limit. 

This cannot be the full story, however. Dimensional reduction of a 6d string theory gives a tower of Kaluza-Klein modes with $\mKK \sim \exp( - 2 \rho /\sqrt{3})$ and a tower of string oscillator modes with $\Ms \sim \exp(  -  \phi/2 -  \rho /(2 \sqrt{3}))$. The two cases we considered above correspond respectively to the limit $\rho \rightarrow \infty$, $\phi$ fixed and the limit $\rho, \phi \rightarrow \infty$, $\rmd \rho / \rmd \phi = 1/\sqrt{3}$ fixed. But this does not include the intermediate regime with $\rho, \phi \rightarrow \infty$, $\rmd \rho / \rmd \phi > 1/\sqrt{3}$ fixed. How are we to see these limits within the 5d vector multiplet moduli space?

The answer is that these limits require curved paths in the space of homogeneous coordinates. Consider once again a path of the form $Y^I(s)$, and let the limit $s \rightarrow 0$ be at infinite distance. By a suitable linear transformation, we may set $Y^I(s=0) = \delta_1^I$. Then, suppose $Y^I(s)$ takes the form: 
\be
Y^1 = 1 \,,~~~Y^2 = s\,,~~~Y^3 = s^\alpha\,,~~~Y^I = 0\,,~~~ I > 3 \,.
\ee
for $0 < \alpha < 1$. With this, the relevant terms in the prepotential $\cF = C_{IJK} Y^I Y^J Y^K / 6$ are those with $I, J, K = 1,2,3$. Since we have assumed that $s \rightarrow 0$ lies at infinite distance, we must have $C_{111} = 0$.
Let us further assume that $C_{333} = C_{133} = C_{113} = C_{112} = 0$: this assumption will prove necessary for our purposes, as we will see momentarily. As a result, the nonzero terms in the prepotential are given by
\begin{align}
\cF= \frac{1}{6} \Big( &C_{222} (Y^2)^3 + 3 C_{122} Y^1 (Y^2)^2  + 6 C_{123} Y^1 Y^2 Y^3 
 + 3 C_{223} (Y^2)^2 Y^3 + 3 C_{233} Y^2 (Y^3)^2  \Big) \,.
\label{Fred}
\end{align}
In the limit $s \rightarrow 0$, the proper distance along the path scales with $s$ as
\begin{equation}
\ell(s) =  \frac{1}{\sqrt{2} } \int_s^1 ds' \sqrt{\mathfrak{g}_{IJ} \dot Y^I \dot Y^J }\,,
\end{equation}
where $\dot{~}$ indicates differentiation with respect to the parameter $s'$, and $g_{IJ}$ is given by \eqref{hgmet}.
Plugging in the prepotential in \eqref{Fred}, we find
\be
\ell(s) = \frac{1}{\sqrt{3}}  \sqrt{\alpha^2 - \alpha + 1} |\log(s)| +...
\ee
where $...$ represents terms that are finite in the limit $s \rightarrow 0$. Meanwhile, the smallest and largest gauge couplings scale with $s$ as 
\begin{align}
g_{\min} &\sim \frac{1}{\sqrt{a_{22}}}  \sim   s^{(2-\alpha)/3} \sim \exp(- \lambda_{\text{KK}} \ell(s) )  \\
 \frac{1}{\sqrt{g_{\max}}} &\sim a_{11}^{1/4} \sim s^{(\alpha+1)/6}  \sim \exp(- \lambda_{\text{string}} \ell(s) )\,,
\end{align}
where
\be
\lambda_{\text{KK}} = \frac{1}{\sqrt{3}} \frac{2 - \alpha}{\sqrt{\alpha^2 - \alpha + 1}}  \,,~~~ \lambda_{\text{string}} = \frac{1}{2 \sqrt{3}} \frac{\alpha + 1}{\sqrt{\alpha^2 - \alpha + 1}}     \,.
\ee
Recall that the tower WGC implies a tower of charged particles beginning at the scale $g_{\text{min}}$, and the WGC for strings implies a tower of string oscillator modes beginning at the scale $1/\sqrt{g_{\max}}$. The masses of these towers therefore decay exponentially in the limit $s \rightarrow 0$ with coefficients $\lambda_{\text{KK}}$ and $\lambda_{\text{string}}$, respectively.

Next, we define an angle $\vartheta$ by
\be
\vartheta = \sin^{-1}\left(\frac{2 - \alpha}{2 \sqrt{\alpha^2 - \alpha + 1}}\right)\,.
\ee
With this, we may rewrite the $\lambda$ coefficients as 
\be
\lambda_{\text{KK}} = \frac{2}{\sqrt{3}}  \sin(\vartheta)  \,,~~~ \lambda_{\text{string}} =  \frac{1}{2} \cos(\vartheta) + \frac{1}{2\sqrt{3}} \sin(\vartheta)     \,.
\ee
This precisely matches the coefficients of \eqref{thetalambda}! We see that the tower implied by the tower WGC matches the expected scaling for a Kaluza-Klein tower, and the string scale implied by the WGC for strings matches the expected scaling for a fundamental string in six dimensions. In other words, we have recovered the ``intermediate regime'' between the strict decompactification limit $\rho \rightarrow \infty$ and the emergent string limit $\rmd \rho / \rmd \phi =  1/\sqrt{3}$.

The crucial ingredient here was the introduction of the parameter $\alpha$. For $\alpha = 0$, we recover the expected scaling behavior for the strict decompactification limit, whereas for $\alpha = 1$ we recover the emergent string limit. As $\alpha$ varies between 0 and 1, we interpolate between these two limits, as shown in Figure \ref{thefig}. Notably, this result did not rely on any top-down input from string theory: the observed scaling behavior of the gauge couplings $g_{\min}$, $g_{\max}$ depends on a delicate interplay between the gauge kinetic matrix $a_{IJ}$ and the metric $g_{IJ}$, but the low-energy 5d supergravity relates these two through the prepotential $\cF$, irrespective of the details of the UV completion.

What happens if we drop the assumption $C_{333} = C_{133} = C_{113} = C_{112} = 0$? If any of these coefficients are nonzero, we find that the scaling above no longer applies. Instead, the presence of any of these terms automatically yields the limiting cases of a strict decompactification or an emergent string limit and renders the intermediate regime $\tan^{-1}(1/\sqrt{3}) < \vartheta < \pi/2$ inaccessible.

In this example, the emergent string limit arises only when both $Y^2$ and $Y^3$ are taken to zero with finite, nonzero $Y^2/Y^3$. In fact, for $h^{1,1} > 2$, emergent string limits always require at least two moduli to vanish. To see this, suppose that only a single modulus vanishes in an emergent string limit, $Y^1 = s$ with all other moduli constant. Since $\cF$ must vanish quadratically in an emergent string limit, this means that $\cF$ must factorize as $\cF = (Y^1)^2 (a_I Y^I)$, where each $a_I$ is c-number coefficient. However, defining $X = a_I Y^I$, we find that that prepotential is given simply by $\cF = (Y^1)^2  X$: in particular, it depends on only two homogeneous coordinates $X$, $Y^1$. This is possible only if $h^{1,1}=2$.

For $h^{1,1} > 2$, therefore, emergent string limits necessarily involve multiple vanishing moduli. This strongly suggests that most emergent string limits arise through the structure we have uncovered above, in which curved paths interpolate smoothly between a strict decompactification limit and an emergent string limit.

\section{Gopakumar-Vafa Invariants and Infinite-Distance Limits}\label{GV}

We have seen that infinite-distance limits in vector multiplet moduli space in 5d supergravity feature the correct scaling of the smallest and largest gauge couplings $g_{\min}$, $g_{\max}$ to correspond to either emergent string limits or decompactification limits to six dimensions. In the latter case, the tower WGC implies a tower of light particles with precisely the correct scaling behavior for a tower of Kaluza-Klein modes.

Towers of Kaluza-Klein modes are not only distinguished by the exponential scaling of their masses with increasing field distance, however; they are also distinguished by an approximately constant density of states. Said differently, the number of particles in a Kaluza-Klein tower of mass $m \approx n \mKK = n/R$ in 5d is approximately independent of $n$.

In simple cases, the number of Kaluza-Klein modes of a given KK momentum $n$ is exactly independent of $n$, as each particle in 6d gives rise to one Kaluza-Klein mode of Kaluza-Klein momentum $n$ for each integer $n$. However, in the presence of discrete symmetries or orbifold actions, certain Kaluza-Klein may be projected out of the spectrum, resulting in a density of states that is periodic rather than constant. One example of this is the ``simple orbifold'' in \S2.2 of \cite{Heidenreich:2016aqi}: in this example, many Kaluza-Klein modes of odd charge are projected out of the spectrum by the orbifold action, and the number of Kaluza-Klein modes of KK momentum $n$ depends only on whether $n$ is odd or even, but it is order-one in either case.

This is in stark contrast to the exponentially large Hagedorn density of string oscillator modes which become light in an emergent string limit; here, the mass of the oscillator mode at level $k$ is given by $m_k \sim \sqrt{k} \Ms$, and the number of oscillator modes at level $k$ scales exponentially with $\sqrt{k}$.\footnote{For the heterotic string in ten dimensions, for example, the degeneracy of oscillator modes at level $n$ scales as $d(k) \sim \exp(2 \pi (2 + \sqrt{2}) \sqrt{k})$ \cite{Heidenreich:2017sim}.} Sometimes, as in the case of heterotic string theory, string oscillator modes may carry gauge charge. In the heterotic case, the charge under a $U(1)$ in the Cartan of the gauge group is quantized, the mass of the lightest state of a given charge $n$ grows linearly with $n$, while its level $k$ scales as $k \sim n^2$. As a result, the number of states with this mass and charge grows exponentially with $\sqrt{k} \sim n$.\footnote{For the heterotic string, this exponential growth comes from the exponentially-large number of choices for the right-moving oscillators needed to level-match the left-moving charge.} We expect that this exponential growth with charge should apply more generally, and we will see that this expectation is borne out in the examples below.

Within the vector multiplet moduli space of a 5d supergravity arising from M-theory on a Calabi-Yau threefold, the spectrum of charged BPS states that become light in an infinite-distance limit are encoded in the Gopakumar-Vafa (GV) invariants of the threefold. By studying how these GV invariants depend on the charge $n$, we can put the Emergent String Conjecture to the test.

In what follows, we will carry out such tests in a large number of Complete Intersection Calabi-Yau threefolds (CICYs). Many of our results follow from the more general analyses of \cite{Lee:2019wij, Cota:2022maf}.

\begin{table}
\begin{center}
$\arraycolsep=5pt
\begin{array}{c|ccccc}
\mathdiagbox[width=1cm,height=0.75cm,innerleftsep=0.1cm,innerrightsep=0cm]{q_2}{q_1}
&0&1&2&3&4 \\ \hline
0&-& 40 & 4& 0&0  \\  
1& \cellcolor{orange!25} 144&496&496&23953120&2388434784 \\ 
2& \cellcolor{orange!25} 164&5616&23100&34528&23100 \\  
3& \cellcolor{orange!25} 144 & 44384 &602016& 2471824 & 4709216 \\  
4& \cellcolor{orange!25} 88 &279976&10439512 &97922024&398659384\\  
5& \cellcolor{orange!25} 144 & 1482384 &136431424& 2616030416&20133562480 \\
6& \cellcolor{orange!25} 164 & 6751472 &1439003864 &52447406096&707697743208 \\
7& \cellcolor{orange!25} 144 &  27208608 & 12779098368 & 841622542048 & 18899196173440 \\
8 &   \cellcolor{orange!25}  88 & 99569856 &  98370714948 &  11277044593704 &  405560003481888 \\
\end{array}
$
\caption{Genus 0 GV invariants of degree $(q_1, q_2)$ for the geometry $X_{7643}$. The BPS particles that become light in the candidate decompactification limit $Y \rightarrow 0$ are shaded orange; these GV invariants are order-one and periodic (of order 4) as a function of $q_2$, consistent with the fact that they represent Kaluza-Klein modes for a circle in another duality frame, which grows to infinite size as $Y \rightarrow 0$.}
\label{firstextab}
\end{center}
\end{table}

As a first example, consider the case of the complete intersection Calabi-Yau $X_{7643}$ with configuration matrix\footnote{For a pedagogical introduction to CICYs and their configuration matrices, see \cite{Hubsch:1992nu}.}
\be
\left(
\begin{array}{c || cccc}
\mathbb{P}^2~ & ~0&0&2&1 \\
\mathbb{P}^5 ~& ~2&2&1&1 \\
\end{array}
\right)
\,.
\ee
In other words, this Calabi-Yau consists of the intersection of four polynomials of bidegree $(0, 2)$, $(0,2)$, $(2, 1)$, and $(1, 1)$, respectively, inside $\mathbb{P}^2 \times \mathbb{P}^5$.
The GV invariants for this geometry can be computed using the Mathematica notebook of Klemm and Kreuzer \cite{Klemm:2001aaa}, which was based on the earlier work \cite{Hosono:1994ax,Hosono:1993qy}, and the result is
shown in Table \ref{firstextab}. 

The mass of a BPS particle of charge $(q_1, q_2)$ is given by 
\be
m_{q_1, q_2} = \left( \frac{\sqrt{2} \pi}{\kappa_5}\right)^{1/3} \frac{|q_1 X + q_2 Y|}{\cF^{1/3}}\,,
\label{BPSmass}
\ee
where $X \equiv Y^1$ measures the volume of the $\mathbb{P}^2$ factor and $Y \equiv Y^2$ measures the volume of the $\mathbb{P}^5$ factor.
In the limit $Y \rightarrow 0$ the tower of particles with $q_1 = 0$, $q_2 = 1, 2, ...$ become massless.
Furthermore, from the form of the prepotential,
\be
\cF = 2 X^2 Y + 6 X Y^2 + \frac{4}{3} Y^3\,,
\ee
we see that the limit $Y \rightarrow 0$ represents a candidate decompactification limit, since the prepotential vanishes linearly in this limit, $\cF \sim Y$. This agrees with the fact that the GV invariants $n_{0, q_2}$ (shaded orange in Table \ref{firstextab}) are order-one and periodic as a function of $q_2$, as expected for a Kaluza-Klein tower. This is especially remarkable when these GV invariants are compared with the GV invariants for $q_1 \neq 0$, which grow exponentially with increasing $q_1$ or $q_2$.

In \cite{Lee:2019wij}, it was shown that such a decompactification limit corresponds geometrically to a limit in which a genus-one fiber shrinks to zero size. The order-one scaling of the GV invariants then follows from modularity of the topological string partition function of genus-one fibrations

\begin{table}
\begin{center}
$\arraycolsep=5pt
\begin{array}{c|ccccc}
\mathdiagbox[width=1cm,height=0.75cm,innerleftsep=0.1cm,innerrightsep=0cm]{q_2}{q_1}
&0&1&2&3&4 \\ \hline
0&-&24&0&0&0 \\  
1& \cellcolor{blue!25}   396 & 1152 & 396 &0 & 0 \\ 
2&\cellcolor{blue!25}  2610&53136& 112068 &53136& 2610 \\  
3& \cellcolor{blue!25} 35640& 2377728& 15951564 &28024704& 15951564 \\  
4& \cellcolor{blue!25}  605844 &103323672& 1602730872 & 6746381496 & 10576809936 \\  
5& \cellcolor{blue!25}  12212172& 4400303616 & 132192153792 & 1084701369600 & 3472953972948 \\
6& \cellcolor{blue!25}  273644244& 184590071136 & 9593083752300 & 135592015659408 & 762494479579314 \\
\end{array}
$
\caption{Genus 0 GV invariants of degree $(q_1, q_2)$ for the geometry $X_{7806}$. The BPS particles that become light in the emergent string limit $Y \rightarrow 0$ are shaded blue; the exponential growth of these GV invariants with $q_2$ is consistent with the exponential, Hagedorn growth of string states with increasing mass.}
\label{secondextab}
\end{center}
\end{table}

As a second example, we consider the case of the CICY $X_{7806}$ with configuration matrix 
\be
\left(
\begin{array}{c || cccc}
\mathbb{P}^1~ & ~0&2 \\
\mathbb{P}^4 ~& ~ 3 &2 \\
\end{array}
\right)
\,.
\ee
In other words, this Calabi-Yau consists of the intersection of two polynomials of bidegree $(0, 3)$ and $(2, 2)$, respectively, inside $\mathbb{P}^1 \times \mathbb{P}^4$.
The GV invariants for this geometry are shown in Table \ref{secondextab}. Here, the mass of a BPS particle of charge $(q_1, q_2)$ is again given by \eqref{BPSmass}, where now $X \equiv Y^1$ measures the size of $\mathbb{P}^1$ and $Y \equiv Y^2$ measures the size of $\mathbb{P}^4$.
In the limit $Y \rightarrow 0$, the tower of particles with $q_1 = 0$, $q_2 = 1, 2, ...$ become massless. Furthermore, from the form of the prepotential,
\be
\cF =  3 X Y^2 + Y^3 \,,
\ee
we see that the limit $Y \rightarrow 0$ represents a candidate emergent string limit, since the prepotential vanishes quadratically in this limit, $\cF \sim Y^2$. This agrees with the fact that the GV invariants $n_{0, q_2}$ (shaded blue in Table \ref{secondextab}) seem to grow exponentially with increasing $q_2$, as expected for a tower in an emergent string limit.\footnote{Throughout this paper, we used a simple eye test to assess exponential growth of the GV invariants, checking that the number of digits in each respective GV invariant grows at a roughly constant rate as charge increases. It would be interesting to measure the exponential rate of growth more precisely and see how it varies across different examples of emergent string limits, but we leave this for future work.}

In \cite{Lee:2019wij}, it was shown that these emergent string limits correspond geometrically to a limit in which a $T^4$ or K3 fiber shrinks to zero size. The former case corresponds to an emergent Type II string, and it is possible that such a limit will have vanishing GV invariants due to enhanced supersymmetry (see e.g. \cite{Cota:2022maf, Gendler:2022ztv}). In the K3 case, however, the GV invariants are nonzero \cite{Lee:2019wij}, and their exponential growth is a consequence of modularity of the topological string partition function \cite{Cota:2022maf}.

\begin{table}
\begin{center}
$\arraycolsep=5pt
\begin{array}{c|ccccc}
\mathdiagbox[width=1cm,height=0.75cm,innerleftsep=0.1cm,innerrightsep=0cm]{q_3}{q_2}
&0&1&2&3&4 \\ \hline
0&-&\cellcolor{blue!25}16&\cellcolor{blue!25}0&\cellcolor{blue!25}0&\cellcolor{blue!25}0 \\  
1& \cellcolor{orange!25}  128 & \cellcolor{blue!25}128 & \cellcolor{blue!25}0 & \cellcolor{blue!25}0 & \cellcolor{blue!25}0  \\ 
2&\cellcolor{orange!25} 144 & \cellcolor{blue!25}960 & \cellcolor{blue!25}144 & \cellcolor{blue!25}0 & \cellcolor{blue!25}0 \\  
3& \cellcolor{orange!25}  128 & \cellcolor{blue!25}5120 & \cellcolor{blue!25}5120 & \cellcolor{blue!25}128 & \cellcolor{blue!25}0 \\  
4& \cellcolor{orange!25} 80 & \cellcolor{blue!25}20640 & \cellcolor{blue!25}70272 & \cellcolor{blue!25}20640 & \cellcolor{blue!25}80 \\  
5& \cellcolor{orange!25} 128 & \cellcolor{blue!25}70656 & \cellcolor{blue!25}626688 & \cellcolor{blue!25}626688 & \cellcolor{blue!25}70656 \\
6& \cellcolor{orange!25} 144 & \cellcolor{blue!25}218752 & \cellcolor{blue!25}4265600 & \cellcolor{blue!25}10349760 & \cellcolor{blue!25}4265600  \\
\end{array}
$
\caption{Genus 0 GV invariants of degree $(0, q_2, q_3)$ for the geometry $X_{7465}$. The BPS particles that become light in the decompactification string limit $Z \rightarrow 0$ are shaded orange, while the additional BPS particles that become light in the emergent string limit $Y, Z \rightarrow 0$ are shaded blue. The GV invariants grow exponentially with increasing charge in the emergent string limit, whereas they are order-one and periodic in the decompactification limit, as expected. }
\label{thirdextab}
\end{center}
\end{table}

As a third and final example, consider the $h^{1,1}=3$ CICY $X_{7465}$ with configuration matrix 
\be
\left(
\begin{array}{c || ccccc}
\mathbb{P}^1~ & ~1 & 1 & 0 & 0 & 0 \\
\mathbb{P}^2 ~& ~ 1 & 0 & 2 & 0 & 0  \\
\mathbb{P}^5 ~& ~ 0 & 1 & 1 & 2 & 2 \\
\end{array}
\right)
\,.
\label{7465ex}
\ee
In other words, this Calabi-Yau consists of the intersection of five polynomials of tridegree $(1,1,1)$, $(1,0,1)$, $(0,2,1)$, $(0,0,2)$, and $(0,0,2)$, respectively, inside $\mathbb{P}^1 \times \mathbb{P}^2 \times \mathbb{P}^5$.\footnote{This geometry is of the type studied in \cite{Cota:2020zse}, which also computed the GV invariants using Noether-Lefschetz theory. Subsequently, \cite{Cota:2022maf} found a match between these and the BPS state counts for the heterotic string, in accordance with expectations.}
The GV invariants with $q_1 = 0$ for this geometry are shown in Table \ref{thirdextab}. Here, the mass of a BPS particle of charge $(q_1, q_2, q_3)$ is given by 
\be
m_{q_1, q_2, q_3} = \left( \frac{\sqrt{2} \pi}{\kappa_5}\right)^{1/3} \frac{|q_1 X + q_2 Y + q_3 Z|}{\cF^{1/3}}\,,
\ee
where $X \equiv Y^1$ measures the volume of the $\mathbb{P}^1$ factor, $Y \equiv Y^2$ measures the volume of the $\mathbb{P}^2$ factor, and $Z \equiv Y^3$ measures the size of the $\mathbb{P}^5$ factor.
Meanwhile, the prepotential is given by
\be
\cF =4 X Y Z + 2 Y^2 Z + 4 X Z^2 + 6 Y Z^2 +   \frac{4}{3} Z^3 \,,
\ee
In the limit $Z \rightarrow 0$, we see that the prepotential vanishes linearly, $\cF \sim Z$. Thus, this corresponds to a candidate decompactification limit. On the other hand, in the limit $Y , Z \rightarrow 0$ with fixed $\rmd Y/ \rmd Z$, the prepotential vanishes quadratically; this represents a candidate emergent string limit. Indeed, this prepotential takes precisely the form \eqref{Fred}, and thus we interpolate between a decompactification limit and an emergent string limit when $X$ is constant, $Z \rightarrow 0$, and $Y \sim Z^\alpha$ for $\alpha \in \{ 0, 1 \}$.

The Emergent String Conjecture suggests that the behavior seen in these particular examples is in fact rather generic. Given any asymptotic boundary in which the prepotential vanishes linearly with some modulus $Y^I$, we expect a periodic tower of order-one GV invariants, which correspond to Kaluza-Klein modes in some duality frame. Meanwhile, any asymptotic boundary in which the prepotential vanishes quadratically with the moduli $Y^I$ corresponds to an emergent string limit, and correspondingly we expect an exponentially-growing tower of charged string modes. If the prepotential vanishes linearly as $Y^I \rightarrow 0$ and quadratically as $Y^I, Y^J \rightarrow 0$, then we expect a periodic tower of order-one GV invariants in the former limit and an exponentially-growing tower of charged string modes in the latter limit.
All this may be viewed as a geometric consequence of the Emergent String Conjecture for Calabi-Yau manifolds.

\section{Numerical Sweep}\label{SWEEP}

Complete intersection Calabi-Yau manifolds in products of projective spaces (CICYs) offer a simple but sizeable arena to test this geometric consequence of the Emergent String Conjecture. In the remainder of this paper, we consider all asymptotic limits of the K\"ahler moduli spaces of CICYs that arise as a single $\mathbb{P}^n$ factor shrinks to zero size. In the language of 5d supergravity, this means that a single homogenous coordinate on vector multiplet moduli space is taken to zero. We further consider all examples of the asymptotic limits discussed in \S\ref{BOTTOMUP}, in which the prepotential takes the form \eqref{Fred}. If the Emergent String Conjecture is true, these limits should interpolate between decompactification limits and emergent string limits depending on the rate at which two of the moduli are taken to zero size.

The GV invariants for all of these geometries can be found in the database of Carta, Mininno, Righi, and Westphal \cite{Carta:2021sms} or, alternatively, computed using the Mathematica code of Klemm and Kreuzer \cite{Klemm:2001aaa}. We examine the GV invariants up to degree 10 to confirm that they exhibit the periodic behavior expected for a decompactification limit and the exponential growth expected for an emergent string limit.

We have seen that decompactification limits are characterized by a prepotential that vanishes linearly with some modulus, $\cF \sim Y^I$, and without loss of generality we take $I=1$. In terms of the triple intersection numbers, this means that $C_{IJK} = 0$ if $I, J, K \neq 1$, but $C_{IJ1} \neq 0$ for some choice of $I, J \neq 1$.

Given a CICY with configuration matrix
\be
\left(
\begin{array}{c || ccccc}
\mathbb{P}^{n_1}~ & ~d_1^{(1)} & d_2^{(1)}  & \cdots & d_k^{(1)} \\
\mathbb{P}^{n_2}~ & ~d_1^{(2)} & d_2^{(2)}  & \cdots & d_k^{(2)} \\
\vdots ~& ~ \vdots&&&\vdots \\
\mathbb{P}^{n_{h^{1,1}}}~ & ~d_1^{(h^{1,1})} & d_2^{(h^{1,1})}  & \cdots & d_k^{(h^{1,1})} \\
\end{array}
\right)
\,,
\label{genconfig}
\ee
the condition $C_{IJK} = 0$ if $I, J, K \neq 1$ is satisfied precisely when the first row $\{ d_j^{(1)} \}$ has fewer than $n_1$ nonzero entries. Meanwhile, the condition $C_{IJ1} \neq 0$ for some choice of $I, J \neq 1$ requires $\{ d_j^{(1)} \}$ to have at least $n_1 -1$ nonzero entries.
The Calabi-Yau condition implies $\sum_j d_j^{(i)} = n_i$, which means that there are only two ways $\{ d_j^{(1)} \}$ can have exactly $n_1-1$ nonzero entries:
\be
\{ d_j^{(1)} \} = \{ 2, 2, \underset{n_1 - 3}{\underbrace{1, ..., 1}}, 0, ..., 0 \} \text{~~~or~~~} \{ d_j^{(1)} \} = \{ 3, \underset{n_1 - 2}{\underbrace{1, ..., 1}}, 0, ..., 0 \} \,,
\ee
or else $\{ d_j^{(1)} \}$ is given by some permutation of the above entries.

We performed a sweep of the favorable CICY database of \cite{Anderson:2017aux} (based on the previous work \cite{Candelas:1987kf}), finding that just 140 of the 7820 favorable CICYs have rows $\{ d_j^{(I)} \}$ of the above form (and only one of these 140, number 7884 in the database, has two rows of this form\footnote{ It is worth noting that CICY number 7884 has $h^{1,1}=2$ and thus a one-dimensional vector multiplet moduli space. The two decompactification limits lie at opposite ends of this moduli space, so they do not intersect, and there is no way to decompactify both circles at once.  }). Computing the GV invariants, we confirm that every row $\{ d_j^{(I)} \}$ of the form corresponds to a periodic tower of order-one GV particles with $q_I = 1, 2, ...$, $q_K = 0$ for $K \neq I$. One such example--number 7643--was discussed in the previous section, and its GV invariants are shown in Table \ref{firstextab}.

Physically speaking, this means that every candidate decompactification limit, for which the prepotential vanishes linearly with the modulus, indeed features a tower of exponentially light particles with an approximately constant density of states, precisely as we would expect for a tower of Kaluza-Klein modes. This gives strong evidence that these limits are indeed decompactification limits in some duality frame, as required by the Emergent String Conjecture.

Another interesting result from our sweep of candidate decompactification limits in CICYs is that the order of periodicity $p$ of the GV invariants Kaluza-Klein tower is always less than or equal to 4. Furthermore, the GV invariants $n_{q}$ of the tower satisfy not only the periodicity condition $n_{q} = n_{q+p}$, but also the condition $n_{q} = n_{p-q}$ for $0 < q < p$. This can be understood as a consequence of charge conjugation, which implies $n_q = n_{-q}$, in conjunction with the periodicity condition.\footnote{We thank Ben Heidenreich for discussions on this point.}

Candidate emergent string limits, on the other hand, have a prepotential which vanishes quadratically with the moduli. To begin, we focus on limits in which a single $\mathbb{P}^{n_I}$ factor shrinks to zero size. Without loss of generality, suppose $I=1$, so that the prepotential vanishes as $\cF \sim (Y^1)^2$. This means that $C_{IJK} = 0$ if $I, J, K \neq 1$, $C_{IJ1} = 0$ for $I, J \neq 1$, but $C_{11K} \neq 0$ for some $K \neq 1$.

For a Calabi-Yau with configuration matrix \eqref{genconfig}, these conditions on the $C_{IJK}$ imply that the first row $\{ d_j^{(1)} \}$ must have precisely $n_1-2$ nonzero entries. There are, once again, only two ways this can happen:
\be
\{ d_j^{(1)} \} = \{ 3, 2, \underset{n_1 - 4}{\underbrace{1, ... , 1}}, 0, ..., 0 \} \text{~~~or~~~} \{ d_j^{(1)} \} = \{ 4, \underset{n_1 - 3}{\underbrace{1, ... , 1}}, 0, ... , 0 \} \,,
\ee
or else $\{ d_j^{(1)} \}$ is given by some permutation of these entries.

A sweep of the CICY database reveals that only 11 of the 7820 favorable CICYs have rows of this form. All of these have $h^{1,1}=2$, as required for an emergent string limit that involves only one vanishing modulus, as shown above. Computing the GV invariants, we find that every row $\{ d_j^{(I)} \}$ of this form corresponds to a tower of exponentially-growing GV invariants with $q_I = 1, 2, ...$, $q_K = 0$ for $K \neq I$. One example of this type--number 7806--was discussed in the previous subsection, and the GV invariants are shown in Table \ref{secondextab}.

Finally, we consider candidate emergent string limits which occur in conjunction with candidate decompactification limits, as in \eqref{7465ex} above, so that a decompactification limit appears when one modulus $Y^1$ is taken to zero while an emergent string limit appears when another modulus $Y^2$ is taken to zero simultaneously. This means, in particular, that the prepotential vanishes linearly as $Y^1 \rightarrow 0$, so $\cF \sim Y^1$, whereas it vanishes quadratically as $Y^1 , Y^2 \rightarrow 0$. Such a prepotential necessarily takes the form in \eqref{Fred}.

Since these emergent string limits necessarily feature decompactification limits and moduli spaces of dimension larger than one, the only geometries that need to be considered in this sweep are the 127 CICYs with candidate decompactification limits and $h^{1,1} > 2$, a subset of the 140 total CICYs with candidate decompactification limits. Given such a small set, it is straightforward to simply compute the triple intersection numbers (and hence the prepotential for every Calabi-Yau) to see if it takes the form in \eqref{Fred}.

We find that of the 127 CICYs with candidate decompactification limits and $h^{1,1} > 2$, 63 also have candidate emergent string limits in which two $\mathbb{P}^{n_I}$ factors shrink to zero size, and all of these have $h^{1,1} = 3$. Furthermore, 40 of the 63 CICYs have two candidate emergent string limits; this means that the prepotential vanishes linearly as $Y^1 \rightarrow 0$, quadratically as $Y^1, Y^2 \rightarrow 0$, and quadratically as $Y^1, Y^3 \rightarrow 0$. In terms of the prepotential in \eqref{Fred}, this happens precisely when $C_{233} = 0$.

Studying the GV invariants of these 63 CICYs, we find once again that every candidate emergent string limit features an exponentially growing tower of BPS particles. One example of this type--number 7465--was discussed in the previous subsection, and the GV invariants are shown in Table \ref{thirdextab}.

Thus, the candidate emergent string limits feature exponential growth of GV invariants, as expected. In summary, our numerical sweep of CICYs provides strong evidence for the Emergent String Conjecture in vector multiplet moduli spaces of 5d supergravity theories

\section{Discussion}\label{CONC}

In this paper, we have seen perfect agreement between the rate of growth of GV invariants along certain rays in the charge lattice and the values of the triple intersection numbers for a large class of Calabi-Yau manifolds. This agreement conforms exactly to the expectations of the Emergent String Conjecture, and thus our work provides strong evidence in favor of this conjecture based on topological invariants of Calabi-Yau manifolds.

To turn this around, our work shows that the Emergent String Conjecture places strong constraints on certain GV invariants, just as previous work has uncovered strong constraints on GV invariants from the tower/sublattice WGC \cite{Alim:2021vhs, Gendler:2022ztv}. Hopefully, the Swampland program will someday yield rigorous predictions that can be tested experimentally, but it is encouraging that it has already given us rigorous, precise, and testable predictions for Calabi-Yau geometry. So far, the Emergent String Conjecture (and the sublattice WGC) have passed these tests with flying colors.

We also found compelling bottom-up evidence for the Emergent String Conjecture by studying curved paths in the vector multiplet moduli spaces of 5d supergravities. Such paths lead to precisely the scaling behavior of gauge couplings for particles and strings that we would expect after circle compactification of a six-dimensional string theory.

A few of our results warrant further study. We saw above many examples of candidate decompactification limits in which the GV invariants are periodic rather than constant with increasing charge. However, in all of the examples we saw, the order of periodicity was less than or equal to 4.
This is reminiscent of the counterexamples to the lattice Weak Gravity Conjecture found in \cite{Heidenreich:2016aqi}; such counterexamples have BPS particles of charge 0 mod $k$ but no BPS particles of charge $q$, $q \neq 0$ mod $k$, and in all examples considered, the order of periodicity satisfies $k \leq 3$.

A key difference between those examples and the examples we have seen in this work is that our examples do not violate the lattice WGC: in Table \ref{firstextab}, for instance, we see that the GV invariants $n_{0, q}$ are all nonzero, indicating that there are BPS particles of all charges $(0, q), q \neq 0$. This is true more generally for all 141 examples of candidate decompactification limits that we have studied, and we find no evidence for violations of the lattice WGC in our examples. However, one intriguing similarity between our results and these is that the periodicity in either case never grows parametrically larger than 1. This is likely also related to the observation that discrete gauge groups in string theory seem to have order-one cardinality (see e.g. \cite{Lee:2022swr}). It would be nice to have a bottom-up argument as to why quantum gravity seems to eschew large periodicities and discrete symmetries.

The CICY database studied here--while sizable--nonetheless represents only a small fraction of all Calabi-Yau manifolds. A logical next step would be to extend the numerical sweep considered here to the full Kreuzer-Skarke database of Calabi-Yau hypersurfaces in ambient toric varieties \cite{Kreuzer:2000xy}.

The bottom-up argument for the Emergent String Conjecture we have given above is suggestive, but it is nonetheless restricted to the case of vector multiplet moduli spaces in 5d supergravity. It would be nice to find a bottom-up argument that doesn't rely on the scaling of gauge couplings, since infinite-distance limits in 5d hypermultiplet moduli spaces do not feature weakly coupled gauge fields, and it would be especially exciting to find an argument for the Emergent String Conjecture that does not rely on supersymmetry at all. We leave these as worthwhile targets for future study.

\section*{Acknowledgements}

It is a pleasure to thank Jonathan Heckman, Ben Heidenreich, and Andrew Turner for useful discussions, Matthew Reece for comments on a draft of this paper, and Timo Weigand and Cesar Fierro Cota for insightful remarks on a preprint of this paper. The author is also indebted to Lara B. Anderson, Xin Gao, James Gray, and Seung-Joo Lee; Albrecht Klemm and Maximilian Kreuzer; Federico Carta, Alessandro Mininno, Nicole Righi, and Alexander Westphal; Philip Candelas, A.M. Dale, Carsten Andrew L\"utken, and Rolf Schimmrigk for their publicly available CICY databases and Mathematica notebooks, which played the starring role in the numerical sweep performed in this work.
This work was supported in part by STFC through grant ST/T000708/1, the Berkeley Center for Theoretical Physics; by the Department of Energy, Office of Science, Office of High Energy Physics under QuantISED Award DE-SC0019380 and under contract DE-AC02-05CH11231; and by the National Science Foundation under Award Number 2112880.

\bibliographystyle{utphys}
\bibliography{ref}
\end{document}